# Harnessing Nonlinear Dynamics for Time-Driven Berry Phase in Classical Systems

Kazi T. Mahmood & M. Arif Hasan*

Department of Mechanical Engineering, Wayne State University, Detroit, MI 48202

kazi.tahsin.mahmood@wayne.edu, hasan.arif@wayne.edu

**Abstract**

Phases arising from cyclic processes are fundamental in physics, bridging quantum and classical domains and providing deeper insights into the topology and dynamics of physical systems. This study investigates the accumulation of a time-driven Berry phase in a classical nonlinear system comprised of two spherical granules and introduces a method in which gauge variants naturally evolve over time without altering internal or external conditions. We develop a perturbation-based model to map the system's elastic characteristics to Bloch states and confirm the theoretical predictions of the frequency-dependent Berry phase through experiments. Our findings reveal that the Berry phase can exhibit trivial and nontrivial values, influenced by external driving forces and static precompression. Our results demonstrate a rich array of vibrational modes, capable of displaying identical Berry phase signatures across different frequencies—a significant departure from previous studies that identified a single topological resonance. Multiple nontrivial Berry phases emerge in highly nonlinear settings, whereas more linear regimes exhibit a singular nontrivial phase. Notably, the behavior of the Berry phase in our system mirrors fundamental quantum mechanics concepts, such as path-dependent state evolution. This study highlights the potential of classical mechanical systems to mimic quantum phenomena, opening new pathways for quantum-inspired topological computation and offering fresh perspectives on using time-driven Berry phase accumulation to investigate topological properties in nonlinear media.

*Corresponding Author

# 1. Introduction

The exploration of geometric and Berry phases has emerged as a pivotal domain in modern physics, bridging the quantum and classical realms through innovative research [1-6]. These phases, which arise during adiabatic cyclic processes, provide deep insights into the dynamics of physical systems. Unlike traditional phases dependent on local properties, geometric phases reflect the global characteristics of a system's topology, thus revealing profound implications for our understanding of fundamental physics [7].

At the heart of this exploration is the Berry phase, a concept that has significantly expanded our understanding of quantum mechanics [8]. Serving as a bridge between abstract theoretical constructs and tangible physical phenomena, the Berry phase illustrates how topological configurations can influence system behavior [9, 10]. Its applications are vast and varied, ranging from enabling decoherence-free operations in quantum computers to explaining unique phenomena in topological insulators and facilitating one-way information flow via the bulk-edge correspondence principle Berry phases enable decoherence-free qubit operations, unique behaviors in topological insulators [11-15]. The global, non-integrable nature of the Berry phase facilitates advanced computational models, such as non-Abelian braiding and Majorana encoding, which are critical for topological quantum computing [16-20].

Historically, the accumulation of the Berry phase has been a focal point in quantum mechanics, particularly in the context of discretized gauge variant quantities like amplitude and phase across different wave numbers [21, 22]. This body of research was also expanded to include time-dependent variants, showing how evolving Hamiltonian parameters can lead to Berry phase accumulation [23, 24].

In the realm of classical mechanics, significant advancements have been made in understanding how geometric phases can accumulate through the dynamics of elastic oscillations, vibrations, and waves within topologically structured systems [3, 4, 25, 26]. These studies have illuminated the role of internal and external parameters in sculpting the Hamiltonian space, providing a foundational understanding of the mechanisms at play. However, such prior studies often focused on static systems where manipulation of internal and external variables was required to guide the evolution of gauge variants [1, 3, 6]. In a significant shift, the present study investigates a novel approach in which gauge variants evolve over time without altering internal system parameters or external drivers, leading to Berry phase accumulation in a classical nonlinear granular system.

Nonlinear granular systems, characterized by discrete interactions between elements, offer a rich laboratory for such investigations due to their responsiveness to external stimuli [27]. These systems, with their intricate dynamics and interparticle forces, present a fertile ground for probing new phenomena that arise from the confluence of nonlinearity, topology, and time-dependent states. A particularly intriguing discovery is the ability of granular systems to mimic quantum-like behaviors such as superposition of states and classical entanglement through elastic bits—an analogue to quantum bits (qubits) [28, 29]. This observation paves the way for a deeper exploration of classical systems exhibiting quantum-like characteristics. In this study, we leverage the nonlinear behavior of granular systems to examine Berry phase accumulation over time. This research establishes a methodology for accumulating the Berry phase through both theoretical and experimental approaches. In Sections 2 and 3, we develop the necessary mathematical modeling to form the elastic bit and the Bloch states, illustrating the accumulation of the Berry phase over time. In Section 4, we outline an experimental procedure to observe the impact of external excitation on the Berry phase. This interdisciplinary approach not only enriches our understanding of topological dynamics in classical mechanics but also provides insight into potential future applications in quantum-inspired computation and information processing.

## 2. Mathematical Modeling of a Driven Nonlinear Granular System: An Analogy to Topological Qubit Dynamics

In this study, we explore the dynamics of a driven nonlinear granular system comprising two spherical granules, drawing parallels to the behavior of Berry phase calculation of a quantum bits (qubits) in quantum-inspired topological computing.

### 2.1. System Description and Mathematical Formulation

Here, we consider two spherical granules, each with mass $m = \frac{4}{3}\pi\rho R^3$, where $R$ is the radius and $\rho$ is the density of the granules, and $u_1$ and $u_2$ are the displacements of the granules from the equilibrium. The system is subjected to external harmonic excitations with magnitudes $F_1$ and $F_2$ at a driving frequency $\omega_D$. The static precompression applied to the granules is denoted by $\delta_0$, and the nonlinear stiffness between the granules, arising from Hertzian contact, is given by $k_{NL} = \frac{E\sqrt{2R}}{3(1-v^2)}$, where $E$ is the Young's modulus and $v$ is the Poisson's ratio of the granules [30]. The damping coefficient $\eta$ models the dissipation in the system. Mathematically, such a nonlinear system can be modeled by the following equations of motion for the displacements $u_1$ and $u_2$ [31]:

$$m\ddot{u}_1 = k_{NL}[F_1 e^{i\omega_D t} - u_1 + \delta_0]_+^{\frac{3}{2}} - k_{NL}[u_1 - u_2 + \delta_0]_+^{\frac{3}{2}} - \eta \dot{u}_1$$

$$m\ddot{u}_2 = k_{NL}[u_1 - u_2 + \delta_0]_+^{\frac{3}{2}} - k_{NL}[u_2 - F_2 e^{i\omega_D t} + \delta_0]_+^{\frac{3}{2}} - \eta \dot{u}_2 \quad (1)$$

Here, $(x)_+$ denotes the positive part of $x$, ensuring that the stiffness terms are non-negative.

To analyze the system's behavior under small displacements, we perform a Taylor series expansion around the static precompression $\delta_0$. Assuming that $\delta_0$ is much larger than the relative displacements of the granules, we expand the nonlinear stiffness terms up to the third order:

$$k_1 = \frac{3}{2}k_{NL}\delta_0^{\frac{1}{2}},\ k_2 = -\frac{3}{8}k_{NL}\delta_0^{-\frac{1}{2}},\ k_3 = -\frac{3}{48}k_{NL}\delta_0^{-\frac{3}{2}}$$

where $k_1 > k_2 > k_3$. This results in the following equations of motion:

$$m\ddot{u}_1 = \left[k_1(F_1 e^{i\omega_D t} - u_1) + k_2(F_1 e^{i\omega_D t} - u_1)^2 + k_3(F_1 e^{i\omega_D t} - u_1)^3 + \cdots\right]$$
$$- [k_1(u_1 - u_2) + k_2(u_1 - u_2)^2 + k_3(u_1 - u_2)^3 + \cdots] - \eta\dot{u}_1$$

$$m\ddot{u}_1 = \left[k_1(F_1 e^{i\omega_D t} - u_1) + k_2(F_1 e^{i\omega_D t} - u_1)^2 + k_3(F_1 e^{i\omega_D t} - u_1)^3 + \cdots\right] \quad (2)$$
$$- [k_1(u_1 - u_2) + k_2(u_1 - u_2)^2 + k_3(u_1 - u_2)^3 + \cdots] - \eta\dot{u}_1$$

Eq. (2) describes a nonlinearly coupled oscillator system, which will lead to quantum like behaviors.

## 3. Theoretical Study

### 3.1. Perturbation Technique and Solution Approach

Given the nonlinear nature of system (2), we employ a regular perturbation technique to obtain approximate solutions [32]. Introducing a small dimensionless parameter $\epsilon$ (with $\epsilon \ll 1$), we rewrite Eq. (2) as follows:

$$m\ddot{u}_1 = \left[k_1(F_1 e^{i\omega_D t} - u_1) + \epsilon k_2(F_1 e^{i\omega_D t} - u_1)^2 + \epsilon k_3(F_1 e^{i\omega_D t} - u_1)^3 + \cdots\right]$$
$$- [k_1(u_1 - u_2) + \epsilon k_2(u_1 - u_2)^2 + \epsilon k_3(u_1 - u_2)^3 + \cdots] - \epsilon\eta\dot{u}_1$$

$$m\ddot{u}_2 = -\left[k_1(u_2 - F_2 e^{i\omega_D t}) + \epsilon k_2(u_2 - F_2 e^{i\omega_D t})^2 + \epsilon k_3(u_2 - F_2 e^{i\omega_D t})^3 + \cdots\right]$$
$$+ [k_1(u_1 - u_2) + \epsilon k_2(u_1 - u_2)^2 + \epsilon k_3(u_1 - u_2)^3 + \cdots] - \epsilon\eta\dot{u}_2$$
(3)

such that the solutions of the displacements can be assumed as

$$u_1 = u_{1,0} + \epsilon u_{1,1} + \epsilon^2 u_{1,2} + \cdots$$
$$u_2 = u_{2,0} + \epsilon u_{2,1} + \epsilon^2 u_{2,2} + \cdots$$

Since the damping ($\eta$) of the system is assumed to be small, we have also introduced $\epsilon$ in the damping term as well in Eq. (3). Here, $u_{1,n-1}, u_{2,n-1}; n = 1,2,3,\ldots$ represents the different orders of the displacements due to the presence of different orders of stiffness, i.e., $k_1, k_2$ and $k_3$. Substituting these expansions into Eq. (3) and equating terms of equal powers of $\epsilon$, we derive a hierarchy of equations for each order:

$\epsilon^0$:
$$m\ddot{u}_{1,0} = k_1[(F_1 e^{i\omega_D t} - u_{1,0}) - (u_{1,0} - u_{2,0})] - \eta\dot{u}_{1,0}$$

$$m\ddot{u}_{2,0} = k_1[(u_{1,0} - u_{2,0}) - (u_{2,0} - F_2 e^{i\omega_D t})] - \eta\dot{u}_{2,0}$$

$\epsilon^1$:
$$m\ddot{u}_{1,1} = k_1[(-u_{1,1}) - (u_{1,1} - u_{2,1})] + k_2\left[(F_1 e^{i\omega_D t} - u_{1,0})^2 - (u_{1,0} - u_{2,0})^2\right]$$
$$- \eta\dot{u}_{1,1}$$

$$m\ddot{u}_{2,1} = k_1[(u_{1,1} - u_{2,1}) - (u_{2,1})] + k_2\left[(u_{1,0} - u_{2,0})^2 - (u_{2,0} - F_2 e^{i\omega_D t})^2\right]$$
$$- \eta\dot{u}_{2,1}$$

$\epsilon^2$:
$$m\ddot{u}_{1,2} = k_1[(-u_{1,2}) - (u_{1,2} - u_{2,2})] + k_2(-2F_1 e^{i\omega_D t} u_{1,1} + 2u_{1,0}u_{1,1})$$
$$- k_2(2u_{1,0}u_{1,1} - 2u_{1,1}u_{2,0} - 2u_{1,0}u_{2,1} + 2u_{2,0}u_{2,1})$$
$$+ k_3\left[(F_1 e^{i\omega_D t} - u_{1,0})^3 - (u_{1,0} - u_{2,0})^3\right] - \eta\dot{u}_{1,2}$$

$$m\ddot{u}_{2,2} = k_1[(u_{1,2} - u_{2,2}) - (u_{2,2})] + k_2(2F_2 e^{i\omega_D t} u_{2,1} - 2u_{2,0}u_{2,1})$$
$$+ k_2(2u_{1,0}u_{1,1} - 2u_{1,1}u_{2,0} - 2u_{1,0}u_{2,1} + 2u_{2,0}u_{2,1})$$
$$+ k_3\left[(u_{1,0} - u_{2,0})^3 - (u_{2,0} - F_2 e^{i\omega_D t})^3\right] - \eta\dot{u}_{2,2}$$
(4)

To solve Eq. (4), we assume the solutions of the form $u_{1,n-1} = nA_{1,n-1}e^{ni\omega_D t}$ and $u_{2,n-1} = nA_{2,n-1}e^{ni\omega_D t}$. Here $A_{1,n-1}$ and $A_{2,n-1}$ represent amplitudes of the granules at the characteristic frequency $n\omega_D$. For example, $A_{1,0}$ and $A_{2,0}$ represent the amplitude response at the driving frequency $\omega_D$, while $A_{1,1}, A_{2,1}, A_{1,2}$ and $A_{2,2}$ represent the higher harmonic amplitude response at the other characteristics frequencies ($2\omega_D, 3\omega_D$) due to the nonlinear term of Eq. (3). After some mathematical steps, these amplitudes can be expressed as (see Supplementary Material Note 1 for the detailed derivation):

$$A_{1,0} = \frac{F_1 k_1 \gamma_1 + F_2 k_1^2}{\gamma_1^2 - k_1^2}, \qquad A_{2,0} = \frac{F_1 k_1^2 + F_2 k_1 \gamma_1}{\gamma_1^2 - k_1^2}$$

$$A_{1,1} = \frac{2k_1 k_2 \sigma_1 + k_2 \gamma_2 \sigma_2}{\gamma_2^2 - 4k_1^2}, \qquad A_{2,1} = \frac{2k_1 k_2 \sigma_2 + k_2 \gamma_2 \sigma_1}{\gamma_2^2 - 4k_1^2}$$

$$A_{1,2} = \frac{12 k_1 k_2 \sigma_5 + 3 k_1 k_3 \sigma_3 + 4 k_2 \gamma_3 \sigma_6 + k_3 \gamma_3 \sigma_4}{\gamma_3^2 - 9k_1^2},$$

$$A_{2,2} = \frac{12 k_1 k_2 \sigma_6 + 3 k_1 k_3 \sigma_4 + 4 k_2 \gamma_3 \sigma_5 + k_3 \gamma_3 \sigma_3}{\gamma_3^2 - 9k_1^2} \tag{5}$$

where, $\gamma_1 = (-m\omega_D^2 + 2k_1 + i\eta\omega_D)$, $\gamma_2 = (-8m\omega_D^2 + 4k_1 + 4i\eta\omega_D)$, $\gamma_3 = (-27m\omega_D^2 + 6k_1 + 9i\eta\omega_D)$, $\sigma_1 = (A_{1,0} - A_{2,0})^2 + (A_{2,0} - F_2)^2$, $\sigma_2 = (F_1 - A_{1,0})^2 - (A_{1,0} - A_{2,0})^2$, $\sigma_3 = (A_{1,0} - A_{2,0})^3 + (A_{2,0} - F_2)^3$, $\sigma_4 = (F_1 - A_{1,0})^3 - (A_{1,0} - A_{2,0})^3$, $\sigma_5 = (F_2 A_{2,1} + A_{1,0} A_{1,1} - A_{1,1} A_{2,0} - A_{1,0} A_{2,1})$, $\sigma_6 = (-F_1 A_{1,1} + A_{1,1} A_{2,0} - A_{1,0} A_{2,1} + A_{2,0} A_{2,1})$.

The above perturbation technique reveals the amplitudes at each of the characteristic's frequency $n\omega_D$ of the nonlinear network of Eq. (3). In the current study, we have limited the perturbation method until the second order of smaller term $\epsilon$, where the zeroth order amplitude represents the linearized term and first and second order amplitudes are the nonlinear response of the system. Yet, the technique can be expanded to include higher-order terms.

### 3.2. Time-Dependent Modal Representation of Bloch States in an Elastic System

The study of classical systems mimicking quantum phenomena presents an exciting frontier in understanding the behavior of complex systems under external influences. In this context, in the previous section, we presented an approach for understanding the dynamics of displacements in a granular system (represented as $u_1$ and $u_2$) under external forces. These displacements were expressed as a sum of linear and nonlinear modes, each characterized by a distinct frequency $n\omega_D$, which we derived using perturbation techniques. The displacements $u_1$ and $u_2$ are approximated as:

$$u_1 = \sum_{n=1}^{3} \epsilon^{n-1} u_{1,n-1} = \sum_{n=1}^{3} \epsilon^{n-1} n A_{1,n-1} e^{ni\omega_D t}, \quad u_2 = \sum_{n=1}^{3} \epsilon^{n-1} u_{2,n-1} = \sum_{n=1}^{3} \epsilon^{n-1} n A_{2,n-1} e^{ni\omega_D t}$$

where $\epsilon$ denotes a small parameter dictating the strength of the nonlinearities. These modes reflect the granular system's response to external excitations at different characteristic frequencies. For each frequency, the displacement field can be expressed in a linear normal mode basis with complex coefficients, leading to the following expression [33]:

$$\begin{pmatrix} nA_{1,n-1} \\ nA_{2,n-1} \end{pmatrix} e^{ni\omega_D t} \equiv (\alpha_n |E_1\rangle + \beta_n |E_2\rangle) e^{ni\omega_D t} \tag{6}$$

Here, $E_1 = \frac{1}{\sqrt{2}} \begin{pmatrix} 1 \\ 1 \end{pmatrix}$ and $E_2 = \frac{1}{\sqrt{2}} \begin{pmatrix} 1 \\ -1 \end{pmatrix}$, represent the in-phase and out-of-phase eigenmodes of the linearized granular system [33, 34]. The coefficients $\alpha_n$ and $\beta_n$ are complex amplitudes corresponding to the contributions of each eigenmode $E_1$ or $E_2$, respectively. These coefficients satisfy the normalization condition $|\alpha_n|^2 + |\beta_n|^2 = 1$. When $\alpha_n = 1$, $\beta_n = 0$, and vice versa, the system is in a pure eigenstate, either $E_1$ or $E_2$. In other cases, a linear superposition of these eigenmodes occurs, forming a superposition state.

While the nonlinear modes do not directly contribute to the orthonormal properties of a system, the combination of $E_1$ and $E_2$ forms a complete orthonormal basis for the granular system. Therefore, using the analogy to quantum mechanics, we apply Dirac notation for the state vectors, as shown in Eq. (6) [33, 34]. This representation enables us to calculate the coefficients $\alpha_n$ and $\beta_n$ from the displacement field (see Supplementary Material Note 2 for the detailed derivation):

$$\alpha_n = \frac{(A_{1,n-1} + A_{2,n-1})}{\sqrt{2}}; \quad \beta_n = \frac{(A_{1,n-1} - A_{2,n-1})}{\sqrt{2}}$$

Using Eq. (5), we get:

$$\alpha_1 = \frac{1}{\sqrt{2}} \frac{k_1(F_1 + F_2)}{(\gamma_1 - k_1)}, \quad \beta_1 = \frac{1}{\sqrt{2}} \frac{k_1(F_1 - F_2)}{(\gamma_1 + k_1)}$$

$$\alpha_2 = \frac{1}{\sqrt{2}} \frac{k_2(\sigma_1 + \sigma_2)}{(\gamma_2 - 2k_1)}, \quad \beta_2 = \frac{1}{\sqrt{2}} \frac{k_2(\sigma_1 - \sigma_2)}{(\gamma_2 + 2k_1)}$$

$$\alpha_3 = \frac{1}{\sqrt{2}} \frac{4k_2(\sigma_5 + \sigma_6) + k_3(\sigma_3 + \sigma_4)}{\gamma_3 - 3k_1}, \quad \beta_3 = \frac{1}{\sqrt{2}} \frac{4k_2(\sigma_6 - \sigma_5) + k_3(\sigma_4 - \sigma_3)}{\gamma_3 + 3k_1} \quad (7)$$

Using these definitions, we can express the total displacement field as a linear superposition of $E_1$ and $E_2$, leading to the following form:

$$\vec{U} = \begin{pmatrix} u_1 \\ u_2 \end{pmatrix} = \sum_{n=1}^{3} \begin{pmatrix} \epsilon^{n-1} n A_{1,n-1} \\ \epsilon^{n-1} n A_{2,n-1} \end{pmatrix} e^{ni\omega_D t}$$
$$\equiv \epsilon^0([\alpha_1|E_1\rangle + \beta_1|E_2\rangle])e^{i\omega_D t} + \epsilon^1([\alpha_2|E_1\rangle + \beta_2|E_2\rangle])e^{2i\omega_D t}$$
$$+ \epsilon^2([\alpha_3|E_1\rangle + \beta_3|E_2\rangle])e^{3i\omega_D t} \equiv [\tilde{\alpha}(t)|E_1\rangle + \tilde{\beta}(t)|E_2\rangle] \quad (8)$$

Here, $\tilde{\alpha}(t) = \left(\sum_{n=1}^{3} \epsilon^{n-1}\alpha_n e^{ni\omega_D t}\right) / \left(\sqrt{|\sum_{n=1}^{3} \epsilon^{n-1}\alpha_n e^{ni\omega_D t}|^2 + |(\sum_{n=1}^{3} \epsilon^{n-1}\beta_n e^{ni\omega_D t})|^2}\right)$ and $\tilde{\beta}(t) = \left(\sum_{n=1}^{3} \epsilon^{n-1}\beta_n e^{ni\omega_D t}\right) / \left(\sqrt{|\sum_{n=1}^{3} \epsilon^{n-1}\alpha_n e^{ni\omega_D t}|^2 + |(\sum_{n=1}^{3} \epsilon^{n-1}\beta_n e^{ni\omega_D t})|^2}\right)$.

The time-dependent coefficients, $\tilde{\alpha}(t)$ and $\tilde{\beta}(t)$ are normalized such that their total magnitude remains unity. Interestingly, while the coefficients $\alpha_n$ and $\beta_n$ are time-independent, the coefficients $\tilde{\alpha}(t)$ and $\tilde{\beta}(t)$ evolve over time, reflecting the coherent nature of the system. These coefficients, as they change in time, determine the relative phase between the states, demonstrating how the granules' vibrational response can exhibit quantum-like behavior, even in a classical system.

This system's behavior mirrors the phenomena observed in quantum mechanics, where the state of a quantum bit or qubit is described as a superposition of two basis states. In quantum mechanics, the state of a qubit is represented in a two-dimensional Hilbert space, using a unit vector that describes a pure state. Similarly, the elastic bit in this classical system can be modeled using a unit vector in a two-dimensional complex vector space (similar to a qubit), representing the superposition of the eigenmodes $E_1$ and $E_2$.

The concept of an elastic bit comes from the idea that, like a qubit, the elastic bit can exist in a superposition of two states. The nonlinearities of the system introduce time-dependent dynamics, allowing the system's state to evolve in time. This time-dependent evolution is crucial because it enables the system to mimic quantum phenomena such as superposition and coherence. Furthermore, by mapping the states onto a Bloch sphere, we can visualize the time evolution of the elastic bit's state.

In quantum mechanics, the state of a qubit is often represented on the Bloch sphere, where the state vector corresponds to a point on the surface of the sphere. Similarly, the state of the elastic bit in this study can be visualized on a Bloch sphere, with two key angles: the polar angle $\tilde{\theta}(t)$ and the azimuthal angle $\tilde{\varphi}(t)$. Using Eq. (8), the modal contribution to the overall displacement field's mode superposition can be expressed as a column displacement state vector $|\psi\rangle$ in terms of $\tilde{\theta}(t)$ and $\tilde{\varphi}(t)$ as such [35] (See Supplementary Material Note 2 for Details):

$$|\psi\rangle = \begin{pmatrix} \cos\dfrac{\tilde{\theta}(t)}{2} \\ e^{i\tilde{\varphi}(t)} \sin\dfrac{\tilde{\theta}(t)}{2} \end{pmatrix} \qquad (9)$$

where

$$\tilde{\theta}(t) = 2\cos^{-1}(|\tilde{\alpha}(t)|), \tilde{\varphi}(t) = \arg(\tilde{\alpha}(t)) - \arg(\tilde{\beta}(t)).$$

These angles, $\tilde{\theta}(t)$ and $\tilde{\varphi}(t)$, describe the superposition of the two eigenmodes $E_1$ and $E_2$ as a time-dependent combination. Hence, the polar angle $\tilde{\theta}(t)$ represents the relative magnitude of the eigenmodes, and the azimuthal angle $\tilde{\varphi}(t)$ encodes the phase difference between them. Hence, the Bloch sphere will provide a geometric representation of the elastic bit state space, with the north and south poles corresponding to the pure states $|E_1\rangle$ and $|E_2\rangle$, respectively. Any other point on the sphere will represent a superposition of these two states. The evolution of these angles on the Bloch sphere will reveal the underlying dynamics of the elastic bit's state, showing periodic oscillations that mirror the quantum behavior of a qubit under external influence.

### 3.3. Manipulating the State Evolution: External Parameters and Nonlinearity

From Eq. (9) we observe that the polar $\tilde{\theta}(t)$ and the azimuthal $\tilde{\varphi}(t)$ angles depend on the complex coefficients $\tilde{\alpha}(t)$ and $\tilde{\beta}(t)$, which are themselves influenced by the external driving forces, frequency, and static precompression. Hence, the evolution of the elastic bit's state is governed by the external driving parameters, including the amplitudes $F_1$ and $F_2$, the frequency $\omega_D$, and the static precompression $\delta_0$. By adjusting these parameters, we can control the path the elastic bit takes on the Bloch sphere. Figure 1 illustrates this by showing the evolution of the elastic bit's state for different driving conditions. The external drivers influence both the polar and azimuthal angles, and variations in these parameters lead to different trajectories on the Bloch sphere.

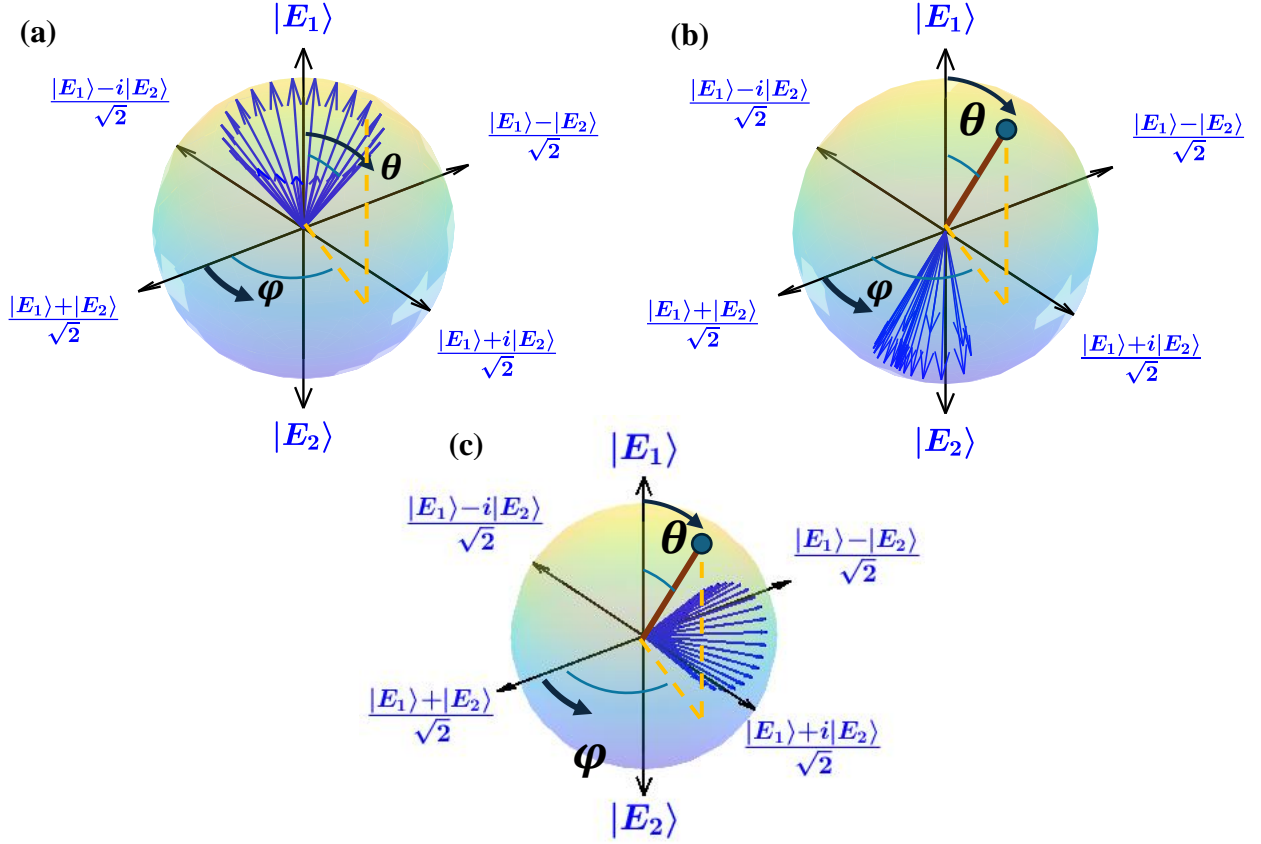

**Figure 1: Time-Dependent State Evolution of Elastic Bits.** The time-dependent evolution of the elastic bit's state, represented on the Bloch sphere, under a specific set of driving frequencies and amplitudes, highlighting the continuous loop and periodic return to the initial state. The polar angle $\tilde{\theta}(t)$ remained nearly constant, indicating that the system's evolution occurred predominantly along the equator of the sphere. Panels (a) to (c) show the effects of varying external stimuli like frequency on the system's varying state evolutions.

Figure 1 shows that the azimuthal angle $\tilde{\varphi}$ makes a complete $2\pi$ revolution in one full period. In contrast, the polar angle $\tilde{\theta}(t)$ variation is negligible, indicating that the state primarily evolves around the equator of the Bloch sphere. As a result, the state of the elastic bit evolves continuously over time. This means that the state vector undergoes periodic motion along the surface of the Bloch sphere, returning to its initial state after each cycle. Here, time permits the parametric exploration of paths on the Bloch sphere without the need to change the external drivers' parameters. In contrast, our previous work has shown the time-independent evolution of the elastic bit states by adjusting the external driver parameters [6].

Furthermore, Fig. 1 illustrates that different evolutions of the states on the Bloch sphere can be achieved by varying external parameters. In Fig. 1, we present three cases in which the elastic bit undergoes rotations around different axes on the Bloch sphere; however, after a certain duration or angle, the state vector returns to its original position. This periodicity is essential for the calculation of the Berry phase, a topological quantity that characterizes the evolution of the state vector over time.

## 3.4. Manipulating the Elastic Bit and Berry Phase Calculation

The Berry phase is a fundamental concept in quantum mechanics that arises when a system undergoes cyclic evolution, and its state vector returns to the same physical state but acquires a phase shift. In this study, we apply the concept of the Berry phase to the elastic bit, providing a quantitative measure of the system's topological properties. By calculating the Berry connection and phase, we gain insight into the underlying geometry of the system's state space and the role of external drivers in shaping its evolution. In the current study, the Berry connection describes how the orientation of the elastic bit's state vector changes as it moves along a path in complex space over time. Summing the Berry connection for possible time values over a closed path gives the Berry phase [10]. Unlike previous studies that defined the Berry connection based on variations in complex amplitude unit vectors tied to wave number [1-3] or external driver parameters [6], our approach focuses on how the state vector changes as a function of time. This approach contrasts with earlier works, which altered the system's topology by modifying internal parameters [1, 2, 4], such as mass or the stiffness of connections, or by changing external parameters [6].

We proceed to establish a framework for which the Berry connection is explicitly defined concerning the time-dependent parameters $\tilde{\theta}(t)$ and $\tilde{\varphi}(t)$. In Eq. (9), we have shown that the elastic bit's state vector $|\psi\rangle$ is expressed in terms of Bloch states $\tilde{\theta}(t)$ and $\tilde{\varphi}(t)$. Based on this, the time-dependent Berry connection can be defined as [36] $BC(t) = i\langle\psi(t)|\partial_t\psi(t)\rangle$; where $|\psi(t)\rangle = \left|\psi\left(\tilde{\theta}(t), \tilde{\varphi}(t)\right)\right\rangle$. To calculate the Berry connection in a time-discretized manner, we divide the time interval into smaller steps of duration $\Delta t$, where $t_n = n\Delta t$ for $n = 1,2,3,...$, and approximate the time evolution over these steps. Applying $\left|\psi\left(\tilde{\theta}(t), \tilde{\varphi}(t)\right)\right\rangle = \begin{pmatrix} \cos\frac{\tilde{\theta}(t)}{2} \\ e^{i\tilde{\varphi}(t)} \sin\frac{\tilde{\theta}(t)}{2} \end{pmatrix}$, we get the overlap along time and, the Berry connection regarding the time-dependent angles can be written as,

$$BC(t) = i\left\langle\psi\left(\tilde{\theta}(t),\tilde{\varphi}(t)\right)\bigg|\psi\left(\tilde{\theta}(t+\Delta t),\tilde{\varphi}(t+\Delta t)\right)\right\rangle$$

$$= i\left[\cos\frac{\tilde{\theta}(t)}{2} \quad e^{-i\tilde{\varphi}(t)}\sin\frac{\tilde{\theta}(t)}{2}\right]\begin{bmatrix}\cos\frac{\tilde{\theta}(t+\Delta t)}{2} \\ e^{i\tilde{\varphi}(t+\Delta t)}\sin\frac{\tilde{\theta}(t+\Delta t)}{2}\end{bmatrix}$$

$$= i\frac{1}{2}\left[\cos\frac{\tilde{\theta}(t)}{2}\cos\frac{\tilde{\theta}(t+\Delta t)}{2} + e^{-i[\tilde{\varphi}(t)-\tilde{\varphi}(t+\Delta t)]}\sin\frac{\tilde{\theta}(t)}{2}\sin\frac{\tilde{\theta}(t+\Delta t)}{2}\right] \quad (10)$$

According to the definition of the Berry phase, it describes the global phase evolution of a complex vector as it is carried around a path in its vector space. So, we have the state vector $|\psi(t)\rangle$, which is a complex quantity, and the phase of the state vector is $Im(\ln(|\psi(t)\rangle))$, where $Im$ takes the imaginary part of its argument. The Berry Phase is then the integral of $BC(t)$ over a closed path in parameter space. We sum up the contributions from each time step around the loop to find the total Berry phase. Since $\langle\psi(t_n)|\psi(t_{n+1})\rangle$ contains phase information about how the state changes between consecutive time steps, the Berry phase $\gamma$ over one loop can be expressed as the argument (imaginary part of the logarithm) of the product of these overlaps in $N$ time steps [11]. If the system undergoes $T_N$ cycles, and a specific Berry phase is accumulated in each cycle where the states follow a specific path. Then the total Berry phase accumulated can be averaged over these $T_N$ cycles. In this case, the expression becomes:

$$\gamma = -\frac{1}{T_N} \text{Im} \sum_{n=1}^{N-1} \ln\langle\psi(t_n)|\psi(t_{n+1})\rangle = -\frac{1}{T_N} \text{Im} \sum_{n=1}^{N-1} \ln[BC(\tilde{\theta}(t_n), \tilde{\varphi}(t_n))] \quad (11)$$

As shown in Eq. (11), the Berry phase is the net phase accumulated over the period that is governed by the external drivers' frequency, and the berry connection is spanned across these response periods. Thus, we can say that the Berry phase characterizes the topology of the elastic bit. Finally, from equations (10) and (11), it is clear that the angles $\theta$ and $\varphi$ influence the Berry phase and Berry connection. Both $\theta$ and $\varphi$, in turn, are affected by the amplitudes, frequency, and phase of the external driver. Therefore, we can manipulate the values of the Berry phase using these parameters, as illustrated in Fig. 2. Finally, from equations (10) and (11), it becomes evident that the angles $\theta$ and $\varphi$ play a significant role in determining the Berry phase and Berry connection. These angular parameters are directly influenced by the characteristics of the external driver, including its amplitudes, frequency, and phase. Consequently, by carefully tuning these external driving parameters, one can effectively control and manipulate the values of the Berry phase. This relationship highlights the intricate interplay between the driving conditions and the geometric properties of the system. Figure 2 clearly illustrates this phenomenon, showcasing how variations in the external driver's parameters translate into changes in the Berry phase, thus offering a practical approach for experimental control in quantum systems.

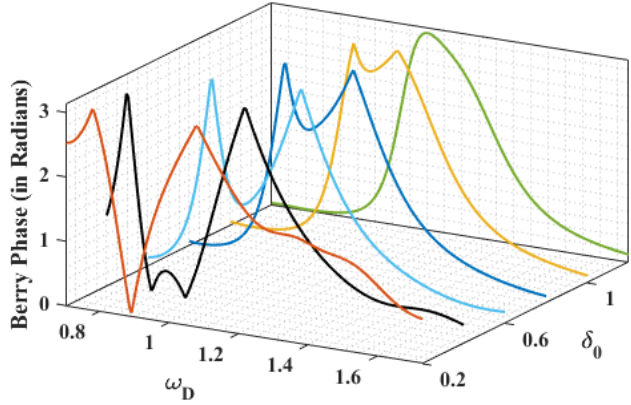

**Figure 2: Frequency and Precompression Dependence of Berry Phase.** Variation of the Berry phase with external driving frequency ($\omega_D$) and static precompression ($\delta_0$). At low precompression, the system exhibits two distinct non-trivial Berry phases of $\pi$, separated by a trivial Berry phase of 0. The shifting behavior of non-trivial Berry phases with increasing static precompression demonstrates a transition from highly nonlinear to weakly nonlinear regimes. The non-trivial phases merge as the system becomes linearized. The plot highlights the sensitivity of the Berry phase to different experimental conditions with varying precompressions. System Parameters: $F_1 = 1$, $F_2 = 2.5$, and $\eta = 0.4$.

Figure 2 demonstrates how Berry phase values vary between 0 and $\pi$ in response to changes in the frequencies ($\omega_D$) of external drivers and the applied static precompression ($\delta_0$). At a lower precompression when the system is nonlinear, there are two instances when the Berry phase attains a non-trivial value of $\pi$, even for a fixed precompression value. Observing two identical non-trivial berry phases indicates the system might have similar vibrational modes at different driving frequency conditions in the nonlinear regime. This contrasts with the findings by Jayaprakash et al., where the authors showed the various families of NNMs and subharmonic orbit modes of the two-granule system in the frequency-energy plot [37]. In Ref. [37], the authors noted families of two NNMs and subharmonic orbits defined over the entire energy range. From Fig. 2, we also observe that at a lower pre-compression value, a trivial berry phase of 0 lies in between the two non-trivial Berry phases. If the value of the static precompression increases, i.e., if we move towards the weakly nonlinear regime, the observance of the trivial berry phase value of 0 vanishes that were observed before in between the non-trivial Berry phases. Further, if the value of the static precompression increases, i.e., if we move towards the weakly nonlinear regime, the external driving frequencies responsible for creating such non-trivial Berry phases shift to higher frequencies. In addition, the external driving frequencies responsible for creating such non-trivial Berry phase values come closer. Hence, the gaps between the external driving frequencies responsible for creating such non-trivial Berry phases shrink. At very high precompression, when the system is close to the linearized regime, the two non-trivial Berry phases merge into one and display one non-trivial Berry phase as we sweep the driving frequency. Thus, Fig. 2 demonstrates the interplay between the extrinsic driver parameters and the intrinsic precompression parameters to achieve the non-trivial Berry phase of $\pi$. Altering the driving frequency allows tuning the static precompression to reach the same non-trivial Berry phase of $\pi$ at different conditions. It is important to note that the occurrence of a non-trivial Berry phase of $\pi$ depends not only on the driving frequency but also on the drivers' amplitude since the system is nonlinear. Such findings highlight the elastic bit's versatility in producing any Berry phase values, including trivial and non-trivial, typically associated with integral multiples of $\pi$.

## 4. Experimental Manifestation of Bloch States and Berry Phase

To experimentally create an elastic bit in a classical nonlinear granular system that can generate a time-dependent superposition of states, we utilize a setup similar to that described in Ref. [33] (see Supplementary Material Note 3 for additional details). It is known that the nonlinearity of the granular system can be controlled by adjusting external driving parameters such as frequency, amplitude, or static precompression [34]. In our theoretical analysis, as shown in Eq. (9), we demonstrate that when the two granules are subjected to static precompression and driven at a specific frequency, the nonlinear vibrational mode can be expressed in a linear normal mode orthonormal basis ($E_1$ and $E_2$) with time-dependent Bloch states ($\tilde{\theta}(t)$ and $\tilde{\varphi}(t)$). These Bloch states constitute the components of a state vector in the two-dimensional Hilbert space of the elastic bit, allowing the system to explore this space parametrically over time along closed periodic paths. Following these theoretical procedures, we use the same orthonormal basis of the elastic bit to map the experimental displacement fields to the Bloch states ($\tilde{\theta}$ and $\tilde{\varphi}$). We then calculate the system's Berry connection and Berry phase using these Bloch states.

Following the theoretical modeling procedures, we used the same orthonormal basis of the elastic bit to map the experimental displacement fields to coherent states. This mapping enables us to track the evolution of the states through the Bloch states ($\tilde{\theta}$ and $\tilde{\varphi}$). We then proceed to calculate the Berry phase of the system using these Bloch states.

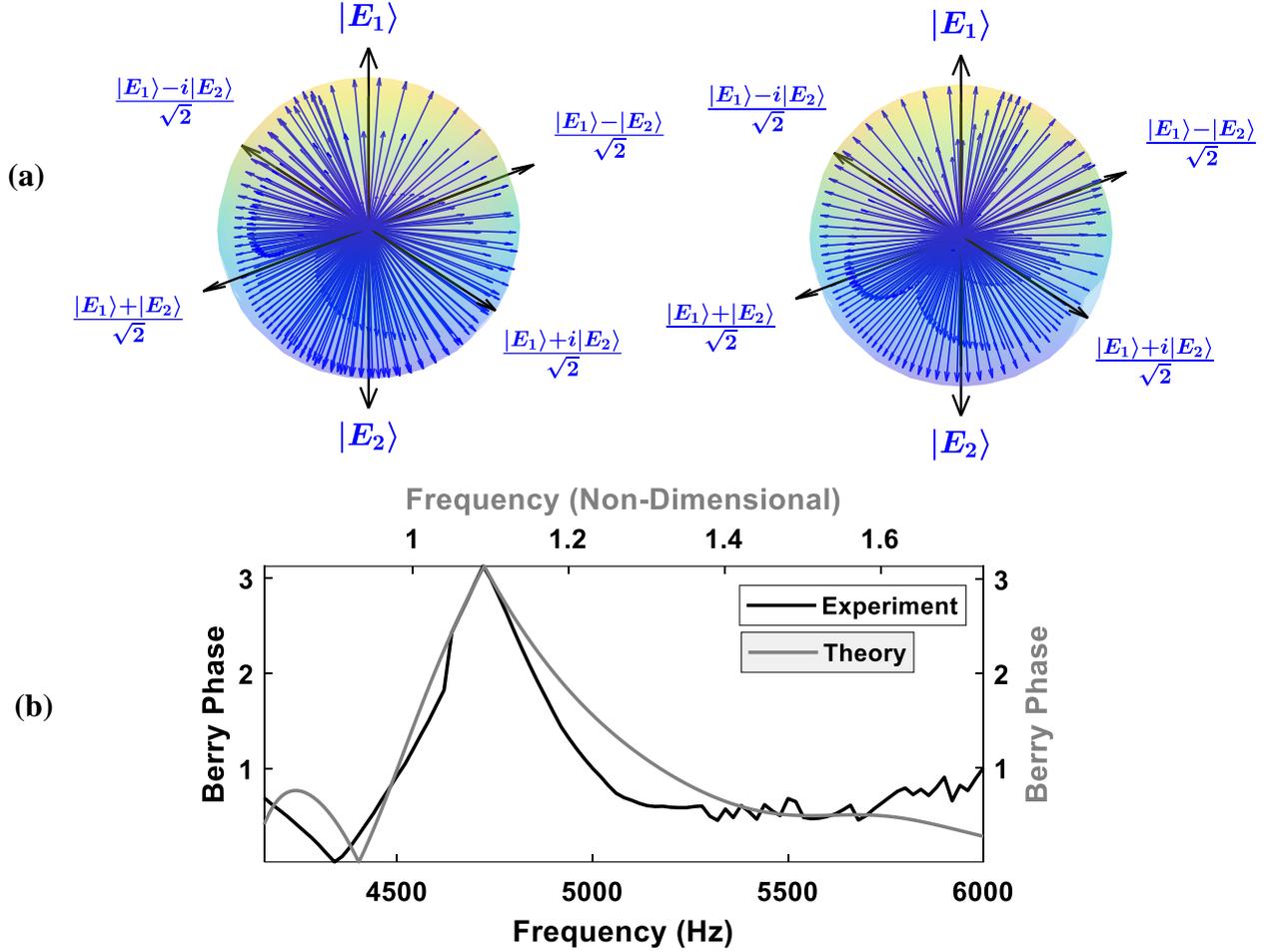

**Figure 3: Experimental Manifestation of Elastic Bit Evolution and Berry Phase Accumulation.** **(a) Trajectories on the Bloch Sphere**: Cyclic evolution of the elastic bit's states on the Bloch sphere at various driving frequencies, illustrating the rotational paths traced by the system's state. The state follows distinct trajectories, highlighting the path dependence of the evolution. The amplitude of the external driver remains constant, while the frequency varies to affect the cyclic path. **(b) Frequency-dependent Berry phase**: variations of Berry phase values in response to changes in driving frequencies, maintaining constant amplitude and static precompression. The ratio of driving frequencies yielding trivial (0) and non-trivial ($\pi$) Berry phases is 1.09 (black line), consistent with the theoretical prediction of 1.18 (gray line).

Figure 3a illustrates the experimental cyclic evolution of the elastic bit's states, represented as rotations on the Bloch sphere at different driving frequencies while keeping the amplitudes of the drivers constant. Compared to Fig. 1, Fig. 3a navigates a broader range of the Bloch sphere and creates twists within a single cyclic evolution. Additionally, Fig. 3a displays distinct trajectories, likely due to the inclusion of higher-order harmonics beyond the third order in the theoretical calculations used for Fig. 1. These higher harmonics introduce unique combinations of amplitude and phases, resulting in these unique trajectories.

Similar to Fig. 1, Fig. 3a clearly shows that the state evolution of the elastic bit is path-dependent. This means that the evolution of the elastic bit is influenced by its final state and the specific path it takes to reach that state. While this is a fundamental property of quantum systems, we demonstrate this behavior in a classical elastic bit system, where the final outcome is determined by both the initial conditions and the trajectory followed. In quantum mechanics, this path dependency leads to phenomena such as interference and the Berry phase. We will show that each unique cyclic evolution of the elastic bit, as illustrated in Fig. 3a, results in a distinct Berry phase. When an elastic bit undergoes a change in a parameter, such as frequency, its state evolves in a way that returns to its original position on the Bloch sphere but gains an additional phase factor in the process. This phase, known as the Berry phase, is determined solely by the elastic bit's path. Figure 3b illustrates how Berry phase values change in response to variations in the frequencies of external drivers while keeping static precompression and driving amplitudes constant. To calculate the Berry connection and Berry phase, we used Eqs. (10) and (11). When the system is driven at a frequency of $\omega_D = 4340$ Hz, a trivial Berry phase of 0 is observed. In contrast, at a frequency of $\omega_D = 4720$ Hz, a non-trivial Berry phase of $\pi$ is recorded. Additionally, intermediate Berry phase values are identified at other frequencies, indicating that the system's response encompasses trivial and non-trivial Berry phase scenarios.

The driving frequency ratio that yields Berry phases of 0 and $\pi$ is calculated to be 4720/4340, which equals 1.09. Comparing this ratio with Fig. 2, we find that the experimental Berry phase results shown in Fig. 3b correlate well with the theoretical plot depicted by the black line in Fig. 2, where $\delta_0$ is set to 0.71. In the theoretical scenario, the ratio of the trivial Berry phase to the non-trivial Berry phase is approximately 1.09/0.92, resulting in about 1.18. This relationship is also illustrated in Fig. 3b for better comparison.

In all cases, the state of an elastic bit evolves along a path that returns to its initial state, although it may accumulate either a zero or a non-zero Berry phase. When the Berry phase is zero, the state has not acquired any additional phase during its evolution; it is effectively the same as the starting point. This scenario is common in quantum mechanics, especially when the evolution is symmetric, or the system's path does not encircle any singularities or topological features in its parameter space.

Conversely, a non-zero Berry phase is more intriguing. It arises when the state's evolution traces a path that encircles a singularity or a topologically significant feature. In this case, while the system ends up in the same physical state (maintaining the same direction on the Bloch sphere), the non-zero value shifts the overall phase of the state. This indicates a phase flip, even though the state is effectively unchanged. This phenomenon is essential for various quantum analogue effects, such as interference and entanglement.

Lastly, the frequency-dependent behavior illustrated in Fig. 3b highlights the nonlinear interactions within the granular system and how they influence Berry phase variations. The ability to experimentally manipulate the Berry phase values demonstrates the potential for using granular systems in topological and non-holonomic analogue computation, where path dependency and external excitation parameters play crucial roles.

## 5. Summary and Concluding Remarks

In this study, we have explored the manifestation of Berry phase accumulation in a nonlinear granular system comprised of two contacting granules, demonstrating how an externally driven elastic bit—a classical analogue of a qubit—exhibits quantum-like state evolution over time. Through theoretical modeling and experimental validation, we have shown how the cyclic evolution of the elastic bit's states on the Bloch sphere results in the accumulation of a nontrivial Berry phase governed by external driving conditions and system nonlinearity.

The investigation began with the development of a mathematical model utilizing a multi-order perturbation method. This model meticulously detailed the amplitudes of each granule at characteristic frequencies, revealing responses at both primary and higher harmonic levels. By employing the linearized normal modes orthonormal basis of the granular system, we successfully mapped the nonlinear complex amplitudes of the displacement field. This mapping allowed us to calculate the time-dependent coefficients of states, which are critical in capturing the dynamics driven by nonlinearity. We have used these coefficients to visually represent the evolution of the elastic bit's state on the Bloch sphere through polar and azimuthal angles. The visualization underscores how the elastic bit evolves and traces a specific path over time, completing a closed loop that allows for the computation of the Berry phase within each cycle. We observed how changes in the excitation frequency could shift the Berry phase from trivial to non-trivial values. Notably, the role of static precompression was pivotal; modifying it allowed transitions from highly to weakly nonlinear behaviors. This adjustment significantly impacted the Berry phase, where highly nonlinear systems displayed multiple non-trivial Berry phase values of $\pi$ across different frequency ranges, while weakly nonlinear systems typically exhibited a singular non-trivial Berry phase as the system approached a linearized state.

Our experimental study provided strong corroboration of these theoretical findings. We constructed a two-mass granular system subjected to external excitation, recording the lateral responses of the granules to determine their time-dependent Bloch states. By applying the same orthonormal basis as in the theoretical framework, we mapped the experimental displacement fields onto coherent states and computed the Berry phase as the system evolved. The experimentally measured frequency-dependent variations in the Berry phase closely aligned with theoretical predictions, with the ratio of trivial to non-trivial Berry phase-driving frequencies calculated as 1.09 in experiments, in good agreement with the theoretical value of 1.18. This aspect of the study underscores the reproducibility and reliability of our theoretical findings in practical scenarios.

Beyond its fundamental significance, the demonstration of Berry phase accumulation in a time-dependent elastic bit system has important implications for topological quantum-inspired computing. The ability to manipulate and control elastic bit states via external parameters opens avenues for implementing fault-tolerant logic operations, similar to topological qubits used in quantum computing. In particular, elastic bit braiding could enable the realization of quantum-like gates such as Hadamard and Phase gates [20]. providing a classical system that mimics quantum coherence and non-Abelian transformations. The observed path-dependent evolution of the elastic bit supports robust, non-local state encoding, reminiscent of Majorana-based quantum computing schemes, where non-local Majorana zero modes exhibit topological resilience to local perturbations [17, 38].

**Acknowledgments**

MAH acknowledges support from NSF grants 2204382 and 2242925.

**Author Contributions**

MAH conceived the idea of the research. KTM fabricated the samples, built the experimental setups, and performed the measurements. All authors analyzed the findings and contributed to the scientific discussion and manuscript writing.

**Competing Interests**

The authors declare no competing interests.

**Data Availability**

The data that support our findings of the present study are available from the corresponding author upon reasonable request.

**Supplementary Materials**

**Note 1: Harmonic Responses of Granules Across Multiple Orders.**

We are investigating a granular system composed of two granules. Considering the one-dimensional motion of the granules, the general mathematical formulation of this two-mass nonlinear granular system is given by Eq. (1). We expand Eq. (1) by applying a power expansion using the Taylor series, yielding the following expression:

$$m\ddot{u}_1 = k_{NL}[F_1 e^{i\omega_D t} - u_1 + \delta_0]_+^{\frac{3}{2}} - k_{NL}[u_1 - u_2 + \delta_0]_+^{\frac{3}{2}} - \eta\dot{u}_1$$

$$\Rightarrow m\ddot{u}_1 = k_{NL}\delta_0^{\frac{3}{2}}\left[1 + \frac{F_1 e^{i\omega_D t} - u_1}{\delta_0}\right]_+^{\frac{3}{2}} - k_{NL}\delta_0^{\frac{3}{2}}\left(1 + \frac{u_1 - u_2}{\delta_0}\right)_+^{\frac{3}{2}} - \eta\dot{u}_1$$

$$\Rightarrow m\ddot{u}_1 = k_{NL}\delta_0^{\frac{3}{2}}\left[1 + \frac{3}{2}\left(\frac{F_1 e^{i\omega_D t} - u_1}{\delta_0}\right) - \frac{3}{8}\left(\frac{F_1 e^{i\omega_D t} - u_1}{\delta_0}\right)^2 - \frac{3}{48}\left(\frac{F_1 e^{i\omega_D t} - u_1}{\delta_0}\right)^3 + \cdots\right]$$

$$- k_{NL}\delta_0^{\frac{3}{2}}\left[1 + \frac{3}{2}\left(\frac{u_1 - u_2}{\delta_0}\right) - \frac{3}{8}\left(\frac{u_1 - u_2}{\delta_0}\right)^2 - \frac{3}{48}\left(\frac{u_1 - u_2}{\delta_0}\right)^3 + \cdots\right] - \eta\dot{u}_1$$

$$\Rightarrow m\ddot{u}_1 = \left[\frac{3}{2}k_{NL}\delta_0^{\frac{1}{2}}(F_1 e^{i\omega_D t} - u_1) - \frac{3}{8}k_{NL}\delta_0^{-\frac{1}{2}}(F_1 e^{i\omega_D t} - u_1)^2 - \frac{3}{48}k_{NL}\delta_0^{-\frac{3}{2}}(F_1 e^{i\omega_D t} - u_1)^3 + \cdots\right]$$

$$- \left[\frac{3}{2}k_{NL}\delta_0^{\frac{1}{2}}(u_1 - u_2) - \frac{3}{8}k_{NL}\delta_0^{-\frac{1}{2}}(u_1 - u_2)^2 - \frac{3}{48}k_{NL}\delta_0^{-\frac{3}{2}}(u_1 - u_2)^3 + \cdots\right] - \eta\dot{u}_1$$

$$m\ddot{u}_2 = -k_{NL}[u_2 - F_2 e^{i\omega_D t} + \delta_0]_+^{\frac{3}{2}} + k_{NL}[u_1 - u_2 + \delta_0]_+^{\frac{3}{2}} - \eta\dot{u}_2$$

$$\Rightarrow m\ddot{u}_2 = -k_{NL}\delta_0^{\frac{3}{2}}\left[1 + \frac{u_2 - F_2 e^{i\omega_D t}}{\delta_0}\right]_+^{\frac{3}{2}} + k_{NL}\delta_0^{\frac{3}{2}}\left(1 + \frac{u_1 - u_2}{\delta_0}\right)_+^{\frac{3}{2}} - \eta\dot{u}_2$$

$$\Rightarrow m\ddot{u}_2 = -k_{NL}\delta_0^{\frac{3}{2}}\left[1 + \frac{3}{2}\left(\frac{u_2 - F_2 e^{i\omega_D t}}{\delta_0}\right) - \frac{3}{8}\left(\frac{u_2 - F_2 e^{i\omega_D t}}{\delta_0}\right)^2 - \frac{3}{48}\left(\frac{u_2 - F_2 e^{i\omega_D t}}{\delta_0}\right)^3 + \cdots\right]$$

$$+ k_{NL}\delta_0^{\frac{3}{2}}\left[1 + \frac{3}{2}\left(\frac{u_1 - u_2}{\delta_0}\right) - \frac{3}{8}\left(\frac{u_1 - u_2}{\delta_0}\right)^2 - \frac{3}{48}\left(\frac{u_1 - u_2}{\delta_0}\right)^3 + \cdots\right] - \eta\dot{u}_2$$

$$\Rightarrow m\ddot{u}_2 = -\left[\frac{3}{2}k_{NL}\delta_0^{\frac{1}{2}}(u_2 - F_2 e^{i\omega_D t}) - \frac{3}{8}k_{NL}\delta_0^{-\frac{1}{2}}(u_2 - F_2 e^{i\omega_D t})^2 - \frac{3}{48}k_{NL}\delta_0^{-\frac{3}{2}}(u_2 - F_2 e^{i\omega_D t})^3 + \cdots\right]$$

$$+ \left[\frac{3}{2}k_{NL}\delta_0^{\frac{1}{2}}(u_1 - u_2) - \frac{3}{8}k_{NL}\delta_0^{-\frac{1}{2}}(u_1 - u_2)^2 - \frac{3}{48}k_{NL}\delta_0^{-\frac{3}{2}}(u_1 - u_2)^3 + \cdots\right] - \eta\dot{u}_2 \quad (A1)$$

By assuming $k_1 = \frac{3}{2}k_{NL}\delta_0^{\frac{1}{2}}$, $k_2 = -\frac{3}{8}k_{NL}\delta_0^{-\frac{1}{2}}$ and $k_3 = -\frac{3}{48}k_{NL}\delta_0^{-\frac{3}{2}}$ leads to Eq. (2). We assume the solution of the displacement field of the granules as $u_1 = u_{1,0} + \epsilon u_{1,1} + \epsilon^2 u_{1,2}$ and $u_2 = u_{2,0} + \epsilon u_{2,1} + \epsilon^2 u_{2,2}$. To carry out the asymptotic method, we introduce a small dimensionless term $\epsilon$.

| | |
|---|---|
| Granule 1 | $m(\ddot{u}_{1,0} + \epsilon \ddot{u}_{1,1} + \epsilon^2 \ddot{u}_{1,2})$<br>$= k_1[(F_1 e^{i\omega_D t} - u_{1,0}) - (u_{1,0} - u_{2,0})]$<br>$+ \epsilon k_1[(-u_{1,1}) - (u_{1,1} - u_{2,1})] + \epsilon^2 k_1[(-u_{1,2}) - (u_{1,2} - u_{2,2})]$<br>$+ \epsilon k_2 \left[ (F_1 e^{i\omega_D t} - u_{1,0})^2 + \epsilon(-2 F_1 e^{i\omega_D t} u_{1,1} + 2 u_{1,0} u_{1,1}) \right.$<br>$- (u_{1,0} - u_{2,0})^2$<br>$\left. - \epsilon^2 (2 u_{1,0} u_{1,1} - 2 u_{1,1} u_{2,0} - 2 u_{1,0} u_{2,1} + 2 u_{2,0} u_{2,1}) \right]$<br>$+ \epsilon^2 k_3 \left[ (F_1 e^{i\omega_D t} - u_{1,0})^3 - (u_{1,0} - u_{2,0})^3 \right]$<br>$- \eta(\dot{u}_{1,0} + \epsilon \dot{u}_{1,1} + \epsilon^2 \dot{u}_{1,2})$ |
| Granule 2 | $m(\ddot{u}_{2,0} + \epsilon \ddot{u}_{2,1} + \epsilon^2 \ddot{u}_{2,2})$<br>$= k_1[(u_{1,0} - u_{2,0}) - (u_{2,0} - F_2 e^{i\omega_D t})]$<br>$+ \epsilon k_1[(u_{1,1} - u_{2,1}) - (u_{2,1})] + \epsilon^2 k_1[(u_{1,2} - u_{2,2}) - (u_{1,2})]$<br>$+ \epsilon k_2 \left[ -(u_{2,0} - F_2 e^{i\omega_D t})^2 + \epsilon(2 F_2 e^{i\omega_D t} u_{2,1} - 2 u_{2,0} u_{2,1}) \right.$<br>$+ (u_{1,0} - u_{2,0})^2$<br>$\left. + \epsilon^2 (2 u_{1,0} u_{1,1} - 2 u_{1,1} u_{2,0} - 2 u_{1,0} u_{2,1} + 2 u_{2,0} u_{2,1}) \right]$<br>$+ \epsilon^2 k_3 \left[ (u_{1,0} - u_{2,0}) - (u_{2,0} - F_2 e^{i\omega_D t})^3 \right]$<br>$- \eta(\dot{u}_{2,0} + \epsilon \dot{u}_{2,1} + \epsilon^2 \dot{u}_{1,2})$     (A2) |

Further simplification of Eq. (A2) by applying $u_1$ and $u_2$ gives us Eq. (3). We can get the multi-order amplitude response by assuming the solution of each of the displacement fields as,

$$\begin{array}{ll}
u_{1,0} = A_{1,0} e^{i\omega_D t} & u_{2,0} = A_{2,0} e^{i\omega_D t} \\
\dot{u}_{1,0} = i\omega_D A_{1,0} e^{i\omega_D t} & \dot{u}_{2,0} = i\omega_D A_{2,0} e^{i\omega_D t} \\
\ddot{u}_{1,0} = -\omega_D^2 A_{1,0} e^{i\omega_D t} & \ddot{u}_{2,0} = -\omega_D^2 A_{2,0} e^{i\omega_D t}
\end{array}$$

$$\begin{array}{ll}
u_{1,1} = 2 A_{1,1} e^{2i\omega_D t} & u_{2,1} = 2 A_{2,1} e^{2i\omega_D t} \\
\dot{u}_{1,1} = 4 i\omega_D A_{1,1} e^{2i\omega_D t} & \dot{u}_{2,1} = 4 i\omega_D A_{2,1} e^{2i\omega_D t} \\
\ddot{u}_{1,1} = -8 \omega_D^2 A_{1,1} e^{2i\omega_D t} & \ddot{u}_{2,1} = -8 \omega_D^2 A_{2,1} e^{2i\omega_D t}
\end{array}$$

$$\begin{array}{ll}
u_{1,2} = 3 A_{1,2} e^{3i\omega_D t} & u_{2,2} = 3 A_{2,2} e^{3i\omega_D t} \\
\dot{u}_{1,2} = 9 i\omega_D A_{1,2} e^{3i\omega_D t} & \dot{u}_{2,2} = 9 i\omega_D A_{2,2} e^{3i\omega_D t} \\
\ddot{u}_{1,2} = -27 \omega_D^2 A_{1,2} e^{3i\omega_D t} & \ddot{u}_{2,2} = -27 \omega_D^2 A_{2,2} e^{3i\omega_D t}
\end{array} \quad (A3)$$

From Eq. (A3), we equate the multi order vibration mode with the order of $\epsilon$ derived in Eq. (4). Using the substitution method, we can get all the amplitude of $A_{1,n-1}$ and $A_{2,n-1}$. At first, the zeroth order amplitudes are calculated as,

$$m\ddot{u}_{1,0} = k_1[(F_1 e^{i\omega_D t} - u_{1,0}) - (u_{1,0} - u_{2,0})] - \eta \dot{u}_{1,0}$$
$$m\ddot{u}_{2,0} = k_1[(u_{1,0} - u_{2,0}) - (u_{2,0} - F_2 e^{i\omega_D t})] - \eta \dot{u}_{2,0}$$

Applying $u_{1,0}$ and $u_{2,0}$ in Eq. (3), we find the simultaneous equation for the amplitude $A_{1,0}$ and $A_{2,0}$ as,

$$-m\omega_D^2 A_{1,0} e^{i\omega_D t} = k_1(F_1 e^{i\omega_D t} - A_{1,0} e^{i\omega_D t}) - k_1(A_{1,0} e^{i\omega_D t} - A_{2,0} e^{i\omega_D t}) - i\eta\omega_D A_{1,0} e^{i\omega_D t}$$

$$\Rightarrow -m\omega_D^2 A_{1,0} e^{i\omega_D t} = [k_1(F_1 - A_{1,0}) - k_1(A_{1,0} - A_{2,0}) - i\eta\omega_D A_{1,0}] e^{i\omega_D t}$$

$$\Rightarrow -m\omega_D^2 A_{1,0} = k_1(F_1 - A_{1,0}) - k_1(A_{1,0} - A_{2,0}) - i\eta\omega_D A_{1,0}$$

$$\Rightarrow (-m\omega_D^2 + 2k_1 + i\eta\omega_D)A_{1,0} - k_1 A_{2,0} = k_1 F_1$$

And

$$-m\omega_D^2 A_{2,0} e^{i\omega_D t} = k_1(A_{1,0} e^{i\omega_D t} - A_{2,0} e^{i\omega_D t}) - k_1(A_{2,0} e^{i\omega_D t} - F_2 e^{i\omega_D t}) - i\eta\omega_D A_{2,0} e^{i\omega_d t}$$

$$\Rightarrow -m\omega_D^2 A_{2,0} e^{i\omega_D t} = [-k_1(A_{2,0} - F_2) + k_1(A_{1,0} - A_{2,0}) - i\eta\omega_D A_{2,0}] e^{i\omega_D t}$$

$$\Rightarrow -m\omega_D^2 A_{1,0} = -k_1(A_{2,0} - F_2) + k_1(A_{1,0} - A_{2,0}) - i\eta\omega_D A_{2,0}$$

$$\Rightarrow -k_1 A_{1,0} + (-m\omega_D^2 + 2k_1 + i\eta\omega_D)A_{2,0} = k_1 F_2$$

We can rewrite the above equations in the compact form as such:

$$\begin{bmatrix} (-m\omega_D^2 + 2k_1 + i\eta\omega_D) & -k_1 \\ -k_1 & (-m\omega_D^2 + 2k_1 + i\eta\omega_D) \end{bmatrix} \begin{pmatrix} A_{1,0} \\ A_{2,0} \end{pmatrix} = \begin{pmatrix} k_1 F_1 \\ k_1 F_2 \end{pmatrix} \quad (A4)$$

Simplifying,

$$\begin{pmatrix} A_{1,0} \\ A_{2,0} \end{pmatrix} = \frac{1}{(-m\omega_D^2 + 2k_1 + i\eta\omega_D)^2 - k_1^2} \begin{bmatrix} (-m\omega_D^2 + 2k_1 + i\eta\omega_D) & k_1 \\ k_1 & (-m\omega_D^2 + 2k_1 + i\eta\omega_D) \end{bmatrix} \begin{pmatrix} k_1 F_1 \\ k_1 F_2 \end{pmatrix}$$

Therefore,

$$A_{1,0} = \frac{F_1 k_1(-m\omega_D^2 + 2k_1 + i\eta\omega_D) + F_2 k_1^2}{(-m\omega_D^2 + 2k_1 + i\eta\omega_D)^2 - k_1^2}$$

$$A_{2,0} = \frac{F_1 k_1^2 + F_2 k_1(-m\omega_D^2 + 2k_1 + i\eta\omega_D)}{(-m\omega_D^2 + 2k_1 + i\eta\omega_D)^2 - k_1^2}$$

For the case of first order amplitude, we get,

$$m\ddot{u}_{1,1} = k_1[(-u_{1,1}) - (u_{1,1} - u_{2,1})] + k_2\left[(F_1 e^{i\omega_D t} - u_{1,0})^2 - (u_{1,0} - u_{2,0})^2\right] - \eta\dot{u}_{1,1}$$

$$m\ddot{u}_{2,1} = k_1[(u_{1,2} - u_{2,2}) - (u_{2,1})] + k_2\left[(u_{1,0} - u_{2,0})^2 - (u_{2,0} - F_2 e^{i\omega_D t})^2\right] - \eta\dot{u}_{2,1}$$

Using substitution method and applying solution of $u_1$ and $u_2$ from Eq. (A3), we get,

$$-8m\omega_D^2 A_{1,1} e^{2i\omega_D t}$$
$$= k_1[(-2A_{1,1} e^{2i\omega_D t}) - (2A_{1,1} e^{2i\omega_D t} - 2A_{2,1} e^{2i\omega_D t})]$$
$$+ k_2\left[(F_1 e^{i\omega_D t} - A_{1,0} e^{i\omega_D t})^2 - (A_{1,0} e^{i\omega_D t} - A_{2,0} e^{i\omega_D t})^2\right]$$
$$- 4i\eta\omega_D A_{1,1} e^{2i\omega_D t}$$

$$\Rightarrow -8m\omega_D^2 A_{1,1} e^{2i\omega_D t}$$
$$= k_1[(-2A_{1,1}) - 2(A_{1,1} - A_{2,1})]e^{2i\omega_D t}$$
$$+ k_2\left[(F_1 - A_{1,0})^2 - (A_{1,0} - A_{2,0})^2\right]e^{2i\omega_D t} - 4i\eta\omega_D A_{1,1} e^{2i\omega_D t}$$

$$\Rightarrow -8m\omega_D^2 A_{1,1}$$
$$= k_1[(-2A_{1,1}) - 2(A_{1,1} - A_{2,1})] + k_2\left[(F_1 - A_{1,0})^2 - (A_{1,0} - A_{2,0})^2\right]$$
$$- 4i\eta\omega_D A_{1,1}$$

$$\Rightarrow (-8m\omega_D^2 + 4k_1 + 4i\eta\omega_D)A_{1,1} - 2k_1 A_{2,1} = k_2\left[(F_1 - A_{1,0})^2 - (A_{1,0} - A_{2,0})^2\right]$$

And

$$-8m\omega_D^2 A_{2,1} e^{2i\omega_D t}$$
$$= k_1[(2A_{1,1}e^{2i\omega_D t} - 2A_{2,1}e^{2i\omega_D t}) - (2A_{2,1}e^{2i\omega_D t})]$$
$$+ k_2\left[(A_{1,0}e^{i\omega_D t} - A_{2,0}e^{i\omega_D t})^2 - (A_{2,0}e^{i\omega_D t} - F_2 e^{i\omega_D t})^2\right]$$
$$- 4i\eta\omega_D A_{2,1} e^{2i\omega_D t}$$

$$\Rightarrow -8m\omega_D^2 A_{1,1} e^{2i\omega_D t}$$
$$= k_1[(2A_{1,1} - 2A_{2,1}) - (2A_{2,1})]e^{2i\omega_D t}$$
$$+ k_2\left[(A_{1,0} - A_{2,0})^2 + (A_{2,0} - F_2)^2\right]e^{2i\omega_D t} - 4i\eta\omega_D A_{2,1} e^{2i\omega_D t}$$

$$\Rightarrow -8m\omega_D^2 A_{2,1}$$
$$= k_1[2(A_{1,1} - A_{2,1}) - (2A_{2,1})] + k_2\left[(A_{1,0} - A_{2,0})^2 + (A_{2,0} - F_2)^2\right]$$
$$- 4i\eta\omega_D A_{2,1}$$

$$\Rightarrow -2k_1 A_{1,1} + (-8m\omega_D^2 + 4k_1 + 4i\eta\omega_D)A_{2,1} = k_2\left[(A_{1,0} - A_{2,0})^2 + (A_{2,0} - F_2)^2\right]$$

Using elimination method, we get,

$$A_{1,1} = \frac{1}{(-8m\omega_D^2 + 4k_1 + 4i\eta\omega_D)^2 - 4k_1^2}\left[2k_1 k_2\left((A_{1,0} - A_{2,0})^2 + (A_{2,0} - F_2)^2\right)\right.$$
$$\left. + k_2\left((F_1 - A_{1,0})^2 - (A_{1,0} - A_{2,0})^2\right)(-8m\omega_D^2 + 4k_1 + 4i\eta\omega_D)\right]$$

$$A_{2,1} = \frac{1}{(-8m\omega_D^2 + 4k_1 + 4i\eta\omega_D)^2 - 4k_1^2}\left[2k_1 k_2\left[(F_1 - A_{1,0})^2 - (A_{1,0} - A_{2,0})^2\right]\right.$$
$$\left. + k_2\left((A_{1,0} - A_{2,0})^2 + (A_{2,0} - F_2)^2\right)(-8m\omega_D^2 + 4k_1 + 4i\eta\omega_D)\right]$$

For the case of second order amplitude, we get,

$$m\ddot{u}_{1,2} = k_1[(-u_{1,2}) - (u_{1,2} - u_{2,2})] + k_2(-2F_1 e^{i\omega_D t}u_{1,1} + 2u_{1,0}u_{1,1})$$
$$- k_2(2u_{1,0}u_{1,1} - 2u_{1,1}u_{2,0} - 2u_{1,0}u_{2,1} + 2u_{2,0}u_{2,1})$$
$$+ k_3\left[(F_1 e^{i\omega_D t} - u_{1,0})^3 - (u_{1,0} - u_{2,0})^3\right] - \eta\dot{u}_{1,2}$$

$$m\ddot{u}_{2,2} = k_1[(u_{1,2} - u_{2,2}) - (u_{2,2})] + k_2(2F_2e^{i\omega_D t}u_{2,1} - 2u_{2,0}u_{2,1})$$
$$+ k_2(2u_{1,0}u_{1,1} - 2u_{1,1}u_{2,0} - 2u_{1,0}u_{2,1} + 2u_{2,0}u_{2,1})$$
$$+ k_3\left[(u_{1,0} - u_{2,0})^3 - (u_{2,0} - F_2e^{i\omega_D t})^3\right] - \eta\dot{u}_{2,2}$$

Substituting $u_1$ and $u_2$ from Eq. (A3), we get,

$$-27m\omega_D^2 A_{1,2}e^{3i\omega_D t}$$
$$= k_1[(-3A_{1,2}e^{3i\omega_D t}) - (3A_{1,2}e^{3i\omega_D t} - 3A_{2,2}e^{3i\omega_D t})]$$
$$+ k_2(-4F_1e^{i\omega_D t}A_{1,1}e^{2i\omega_D t} + 4A_{1,0}e^{i\omega_D t}A_{1,1}e^{2i\omega_D t})$$
$$- k_2(4A_{1,0}e^{i\omega_D t}A_{1,1}e^{2i\omega_D t} - 4A_{1,1}e^{2i\omega_D t}A_{1,0}e^{i\omega_D t} - 4A_{1,0}e^{i\omega_D t}A_{2,1}e^{2i\omega_D t}$$
$$+ 4A_{2,0}e^{i\omega_D t}A_{2,1}e^{2i\omega_D t})$$
$$+ k_3\left[(F_1e^{i\omega_D t} - A_{1,0}e^{i\omega_D t})^3 - (A_{1,0}e^{i\omega_D t} - A_{2,0}e^{i\omega_D t})^3\right]$$
$$- 9i\eta\omega_D A_{1,2}e^{3i\omega_D t}$$

$$\Rightarrow -27m\omega_D^2 A_{1,2}e^{3i\omega_D t}$$
$$= k_1[(-3A_{1,2}) - 3(A_{1,2} - A_{2,2})]e^{3i\omega_D t} + 4k_2(-F_1A_{1,1} + A_{1,0}A_{1,1})e^{3i\omega_D t}$$
$$- 4k_2(A_{1,0}A_{1,1} - A_{1,1}A_{1,0} - A_{1,0}A_{2,1} + A_{2,0}A_{2,1})e^{3i\omega_D t}$$
$$+ k_3\left[(F_1 - A_{1,0})^3 - (A_{1,0} - A_{2,0})^3\right]e^{3i\omega_D t} - 9i\eta\omega_D A_{1,2}e^{3i\omega_D t}$$

$$\Rightarrow -27m\omega_D^2 A_{1,2}$$
$$= k_1[(-3A_{1,2}) - 3(A_{1,2} - A_{2,2})]$$
$$+ 4k_2(-F_1A_{1,1} - A_{1,1}A_{1,0} - A_{1,0}A_{2,1} + A_{2,0}A_{2,1})$$
$$+ k_3\left[(F_1 - A_{1,0})^3 - (A_{1,0} - A_{2,0})^3\right] - 9i\eta\omega_D A_{1,2}$$

$$\Rightarrow (-27m\omega_D^2 + 6k_1 + 9i\eta\omega_D)A_{1,2} - 3k_1A_{2,2}$$
$$= 4k_2(-F_1A_{1,1} - A_{1,1}A_{1,0} - A_{1,0}A_{2,1} + A_{2,0}A_{2,1})$$
$$+ k_3\left[(F_1 - A_{1,0})^3 - (A_{1,0} - A_{2,0})^3\right]$$

And,

$$-27m\omega_D^2 A_{2,2}e^{3i\omega_D t}$$
$$= k_1[(3A_{1,2}e^{3i\omega_D t} - 3A_{2,2}e^{3i\omega_D t}) - (3A_{2,2}e^{3i\omega_D t})]$$
$$+ k_2(4F_2e^{i\omega_D t}A_{2,1}e^{2i\omega_D t} - 4A_{2,0}e^{i\omega_D t}A_{2,1}e^{2i\omega_D t})$$
$$+ k_2(4A_{1,0}e^{i\omega_D t}A_{1,1}e^{2i\omega_D t} - 4A_{1,1}e^{2i\omega_D t}A_{2,0}e^{i\omega_D t} - 4A_{1,0}e^{i\omega_D t}A_{2,1}e^{2i\omega_D t}$$
$$+ 4A_{2,0}e^{i\omega_D t}A_{2,1}e^{2i\omega_D t})$$
$$+ k_3\left[(A_{1,0}e^{i\omega_D t} - A_{2,0}e^{i\omega_D t})^3 + (A_{2,0}e^{i\omega_D t} - F_2e^{i\omega_D t})^3\right]$$
$$- 9i\eta\omega_D A_{2,2}e^{3i\omega_D t}$$

$$\Rightarrow -27m\omega_D^2 A_{2,2}e^{3i\omega_D t}$$
$$= k_1[3(A_{1,2} - A_{2,2}) - (3A_{2,2})]e^{3i\omega_D t} + 4k_2(F_2A_{2,1} - A_{2,0}A_{2,1})e^{3i\omega_D t}$$
$$+ 4k_2(A_{1,0}A_{1,1} - A_{1,1}A_{2,0} - A_{1,0}A_{2,1} + A_{2,0}A_{2,1})e^{3i\omega_D t}$$
$$+ k_3\left[(A_{1,0} - A_{2,0})^3 + (A_{2,0} - F_2)^3\right]e^{3i\omega_D t} - 9i\eta\omega_D A_{2,2}e^{3i\omega_D t}$$

$$\Rightarrow -27m\omega_D^2 A_{2,2}$$
$$= k_1[3(A_{1,2} - A_{2,2}) - (3A_{2,2})]$$
$$+ 4k_2(F_2 A_{2,1} + A_{1,0}A_{1,1} - A_{1,1}A_{2,0} - A_{1,0}A_{2,1})$$
$$+ k_3\left[(A_{1,0} - A_{2,0})^3 + (A_{2,0} - F_2)^3\right] - 9i\eta\omega_D A_{2,2}$$

$$\Rightarrow -3k_1 A_{1,2} + (-27m\omega_D^2 + 6k_1 + 9i\eta\omega_D)A_{2,2}$$
$$= 4k_2(F_2 A_{2,1} + A_{1,0}A_{1,1} - A_{1,1}A_{2,0} - A_{1,0}A_{2,1})$$
$$+ k_3\left[(A_{1,0} - A_{2,0})^3 + (A_{2,0} - F_2)^3\right]$$

Using elimination method, we get,

$$A_{1,2} = \frac{1}{(-27m\omega_D^2 + 6k_1 + 9i\eta\omega_D)^2 - 9k_1^2}\left[12k_1 k_2(F_2 A_{2,1} + A_{1,0}A_{1,1} - A_{1,1}A_{2,0} - A_{1,0}A_{2,1})\right.$$
$$+ 3k_1 k_3\left((A_{1,0} - A_{2,0})^3 - (A_{2,0} - F_2)^3\right)$$
$$+ \left(4k_2(-F_1 A_{1,1} + A_{1,1}A_{2,0} - A_{1,0}A_{2,1} + A_{2,0}A_{2,1})\right.$$
$$\left.\left. + k_3\left((F_1 - A_{1,0})^3 - (A_{1,0} - A_{2,0})^3\right)\right)(-27m\omega_D^2 + 6k_1 + 9i\eta\omega_D)\right]$$

$$A_{2,2} = \frac{1}{(-27m\omega_D^2 + 6k_1 + 9i\eta\omega_D)^2 - 9k_1^2}\left[12k_1 k_2(-F_1 A_{1,1} + A_{1,1}A_{2,0} - A_{1,0}A_{2,1} + A_{2,0}A_{2,1})\right.$$
$$+ 3k_1 k_3\left((F_1 - A_{1,0})^3 - (A_{1,0} - A_{2,0})^3\right)$$
$$+ \left(4k_2(F_2 A_{2,1} + A_{1,0}A_{1,1} - A_{1,1}A_{2,0} - A_{1,0}A_{2,1})\right.$$
$$\left.\left. + k_3\left((A_{1,0} - A_{2,0})^3 + (A_{2,0} - F_2)^3\right)\right)(-27m\omega_D^2 + 6k_1 + 9i\eta\omega_D)\right]$$

We can replace the common terms with $\gamma_1, \gamma_2, \gamma_3, \sigma_1, \sigma_2, \sigma_3, \sigma_4, \sigma_5, \sigma_6$, which gives us the amplitude as in Eq. (5).

**Note 2: Representation of the Coefficients for the Superposition of States**

To implement the orthonormal basis of the total displacement field $\vec{U}$, we get the linear eigenstates from the zeroth order displacement response of the granular network [39-41]. From Eq. (A4), by taking the real part of the stiffness matrix $K$, the eigenmode frequencies is calculated as $\det(K) = 0$, which gives us the eigenmode frequencies as, $\omega_{01}^2 = \sqrt{\frac{k_1}{m}}$ and $\omega_{02}^2 = \sqrt{\frac{3k_1}{m}}$. To get the eigenstates, we apply the eigenmode frequencies in Eq. (A4),

When $\omega_D = \omega_{01} = \sqrt{\frac{k_1}{m}}$,

$$\begin{bmatrix} -m\frac{k_1}{m} + 2k_1 & -k_1 \\ -k_1 & -m\frac{k_1}{m} + 2k_1 \end{bmatrix} \begin{pmatrix} A_{1,0} \\ A_{2,0} \end{pmatrix} = 0 \Rightarrow \begin{bmatrix} k_1 & -k_1 \\ -k_1 & k_1 \end{bmatrix} \begin{pmatrix} A_{1,0} \\ A_{2,0} \end{pmatrix} = 0$$

$$\Rightarrow k_1 \begin{bmatrix} 1 & -1 \\ -1 & 1 \end{bmatrix} \begin{pmatrix} A_{1,0} \\ A_{2,0} \end{pmatrix} = 0$$

Both equations yield the same results, namely, $A_{1,0} = A_{2,0}$. So, $\begin{pmatrix} A_{1,0} \\ A_{2,0} \end{pmatrix}$ is proportional to vector $\begin{pmatrix} 1 \\ 1 \end{pmatrix}$. After normalization, we get the eigenstate corresponding to $\omega_{01}$ as $E_1 = \frac{1}{\sqrt{2}}\begin{pmatrix} 1 \\ 1 \end{pmatrix}$.

When $\omega_D = \omega_{01} = \sqrt{\frac{3k_1}{m}}$,

$$\begin{bmatrix} -m\frac{3k_1}{m} + 2k_1 & -k_1 \\ -k_1 & -m\frac{3k_1}{m} + 2k_1 \end{bmatrix} \begin{pmatrix} A_{1,0} \\ A_{2,0} \end{pmatrix} = 0 \Rightarrow \begin{bmatrix} -k_1 & -k_1 \\ -k_1 & -k_1 \end{bmatrix} \begin{pmatrix} A_{1,0} \\ A_{2,0} \end{pmatrix} = 0$$

$$\Rightarrow k_1 \begin{bmatrix} -1 & -1 \\ -1 & -1 \end{bmatrix} \begin{pmatrix} A_{1,0} \\ A_{2,0} \end{pmatrix} = 0$$

So, the equations give us $A_{1,0} = -A_{2,0}$. So, $\begin{pmatrix} A_{1,0} \\ A_{2,0} \end{pmatrix}$ is proportional to vector $\begin{pmatrix} 1 \\ -1 \end{pmatrix}$. After normalization, we get the eigenstate corresponding to $\omega_{02}$ as $E_1 = \frac{1}{\sqrt{2}}\begin{pmatrix} 1 \\ -1 \end{pmatrix}$. This $E_1$ and $E_2$ demonstrates the eigenstates of a linearized granular network [6]. These are the corresponding in-phase and out-of-phase modes of the system.

Using the different orders of amplitudes $(A_{1,n-1}$ and $A_{2,n-1})$ and the eigenstates of the granules $(E_1 = \frac{1}{\sqrt{2}}\begin{pmatrix} 1 \\ 1 \end{pmatrix}$ and $E_2 = \frac{1}{\sqrt{2}}\begin{pmatrix} 1 \\ -1 \end{pmatrix})$, we express the amplitude of the each order based on the characteristics frequency $n\omega_n$, through the superposition of states' coefficients,

$$\begin{pmatrix} nA_{1,n-1} \\ nA_{2,n-1} \end{pmatrix} e^{ni\omega_D t} \equiv \frac{1}{\sqrt{|\alpha_n|^2 + |\beta_n|^2}} (\alpha_n E_1 + \beta_n E_2) e^{ni\omega_n t} \tag{A5}$$

Here, $\alpha_n$ and $\beta_n$ are the complex amplitude coefficients stating the orthonormal basis at the characteristic frequency of the displacement field. And $n = 1,2,3,....$ are the $n^{th}$ complex amplitudes of the $n^{th}$ dominant characteristic frequency. $\alpha_n$ and $\beta_n$ are dependent on each other through phase and form a coherent superposition of states in the form of two possible vibration modes. We can write the complex amplitude coefficients based on the amplitude of the granules $A_{1,n-1}$ and $A_{2,n-1}$ for each of the characteristics frequency $\omega_n$. We use the analogy of the quantum system, Dirac notations for vectors, and apply them to the elastic states in Eq. (A5).

$$\begin{pmatrix} nA_{1,n-1} \\ nA_{2,n-1} \end{pmatrix} = [\alpha_n|E_1\rangle + \beta|E_2\rangle] = \left[\frac{1}{\sqrt{2}}\alpha_n \begin{pmatrix} 1 \\ 1 \end{pmatrix} + \frac{1}{\sqrt{2}}\beta_n \begin{pmatrix} 1 \\ -1 \end{pmatrix}\right] = \frac{1}{\sqrt{2}}\begin{bmatrix} \alpha_n + \beta_n \\ \alpha_n - \beta_n \end{bmatrix} \quad (A6)$$

We can express the coefficients through the amplitude of the granules. Each of the characteristic frequencies have coefficients that are dependent on the material property and external force. From here, we get $\alpha_n = \frac{(A_{1,n-1}+A_{2,n-1})}{\sqrt{2}}$ and $\beta_n = \frac{(A_{1,n-1}-A_{2,n-1})}{\sqrt{2}}$. For each of the characteristics frequency, we can get the complex amplitude coefficients by calculating the $\alpha_n$ and $\beta_n$ through the amplitude of the granules, we get all the coefficient at each of the characteristics Eq. as (7).

Hence, the total displacement field has a complex time-dependent coefficient. $\tilde{\alpha}(t)$ and $\tilde{\beta}(t)$, which can be expressed as,

$$\vec{U} = \epsilon^0 \begin{pmatrix} A_{1,0} \\ A_{2,0} \end{pmatrix} e^{i\omega_D t} + \epsilon^1 \begin{pmatrix} 2A_{1,1} \\ 2A_{2,1} \end{pmatrix} e^{2i\omega_D t} + \epsilon^2 \begin{pmatrix} 3A_{1,2} \\ 3A_{2,2} \end{pmatrix} e^{3i\omega_D t}$$
$$\equiv \epsilon^0(\alpha_1|E_1\rangle + \beta_1|E_2\rangle)e^{i\omega_D t} + \epsilon^1(\alpha_2|E_1\rangle + \beta_2|E_2\rangle)e^{2i\omega_D t}$$
$$+ \epsilon^2(\alpha_3|E_1\rangle + \beta_3|E_2\rangle)e^{3i\omega_D t}$$

The time dependent coefficients can be calculated as,

$$\Rightarrow \vec{U} = \sum_{n=1}^{3} \begin{pmatrix} \epsilon^{n-1}nA_{1,n-1} \\ \epsilon^{n-1}nA_{2,n-1} \end{pmatrix} e^{ni\omega_D t}$$
$$\equiv \epsilon^0(\alpha_1 e^{i\omega_D t}|E_1\rangle + \beta_1 e^{i\omega_D t}|E_2\rangle) + \epsilon^1(\alpha_2 e^{2i\omega_D t}|E_1\rangle + \beta_2 e^{2i\omega_D t}|E_2\rangle)$$
$$+ \epsilon^2(\alpha_3 e^{3i\omega_D t}|E_1\rangle + \beta_3 e^{3i\omega_D t}|E_2\rangle)$$

$$\Rightarrow \vec{U} = \sum_{n=1}^{3} \begin{pmatrix} \epsilon^{n-1}nA_{1,n-1} \\ \epsilon^{n-1}nA_{2,n-1} \end{pmatrix} e^{ni\omega_D t}$$
$$\equiv (\tilde{\alpha}_1(t)|E_1\rangle + \tilde{\beta}_1(t)|E_2\rangle) + (\tilde{\alpha}_2(t)|E_1\rangle + \tilde{\beta}_2(t)|E_2\rangle)$$
$$+ (\tilde{\alpha}_3(t)|E_1\rangle + \tilde{\beta}_3(t)|E_2\rangle)$$

$$\Rightarrow \vec{U} = \sum_{n=1}^{3} \begin{pmatrix} \epsilon^{n-1}nA_{1,n-1} \\ \epsilon^{n-1}nA_{2,n-1} \end{pmatrix} e^{ni\omega_D t}$$
$$\equiv [\tilde{\alpha}_1(t) + \tilde{\alpha}_2(t) + \tilde{\alpha}_3(t)]|E_1\rangle + [\tilde{\beta}_1(t) + \tilde{\beta}_2(t) + \tilde{\beta}_3(t)]|E_2\rangle$$

$$\Rightarrow \vec{U} = \sum_{n=1}^{3} \begin{pmatrix} \epsilon^{n-1}nA_{1,n-1} \\ \epsilon^{n-1}nA_{2,n-1} \end{pmatrix} e^{ni\omega_D t} \equiv \frac{1}{\sqrt{|\tilde{\alpha}(t)|^2 + |\tilde{\beta}(t)|^2}}[\tilde{\alpha}(t)|E_1\rangle + \tilde{\beta}(t)|E_2\rangle]$$

This allows the coherent states coefficient to become time-dependent, as the multiple characteristics frequency allows time to evolve in the Hilbert space. We can express the coefficients of $\tilde{\alpha}(t)$ and $\tilde{\beta}(t)$ in terms of the amplitude of the granules $A_{1,n-1}$ and $A_{2,n-1}$ as,

$$\tilde{\alpha}(t) = \left(\sum_{n=1}^{3} \frac{1}{\sqrt{|\alpha_n|^2+|\beta_n|^2}} \epsilon^{n-1} \alpha_n e^{ni\omega_n t}\right) = \sum_{n=1}^{3} \frac{1}{\sqrt{2\left(|A_{1,n-1}|^2+|A_{2,n-1}|^2\right)}} (A_{1,n-1} + A_{2,n-1}) \epsilon^{n-1} e^{ni\omega_n t}$$

and

$$\tilde{\beta}(t) = \left(\sum_{n=1}^{3} \frac{1}{\sqrt{|\alpha_n|^2+|\beta_n|^2}} \epsilon^{n-1} \beta_n e^{ni\omega_n t}\right) = \sum_{n=1}^{3} \frac{1}{\sqrt{2\left(|A_{1,n-1}|^2+|A_{2,n-1}|^2\right)}} (A_{1,n-1} - A_{2,n-1}) \epsilon^{n-1} e^{ni\omega_n t}.$$

The modal representation of the superposition of states' coefficients can be written as,

$$|\psi\rangle = \begin{pmatrix} \tilde{\alpha}(t) \\ \tilde{\beta}(t) \end{pmatrix} \tag{A7}$$

Implying the coefficients $\tilde{\alpha}(t)$ and $\tilde{\beta}(t)$ in the displacement field, we can apply a global phase that does not alter states' superposition. With that, we can divide the coefficients into angular quantities, which are the Bloch states' polar $\left(\tilde{\theta}(t)\right)$ and azimuthal $\left(\tilde{\varphi}(t)\right)$ [6, 35, 42]. These angles represent the position and evolution of the elastic bit in the Bloch sphere over time.

**Note 3: Procedure for Conducting the Experiment**

We develop an experimental setup to observe the Bloch states' evolution and the Berry phase accumulation. Fig. A1 shows a driven nonlinear granular network, which creates an elastic bit.

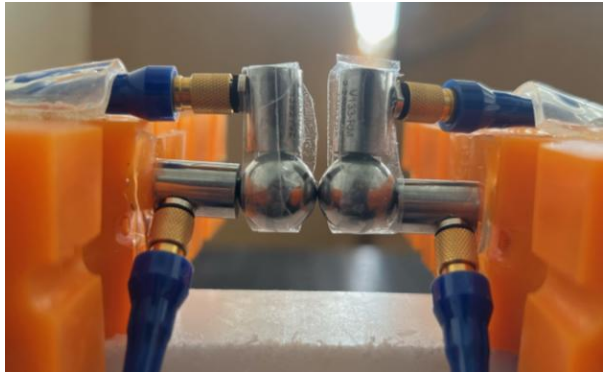

**Figure A1:** Experimental setup of the nonlinear granular network. One of the granules is driven by an ultrasonic transducer. Meanwhile, the lateral transducers record the responses of the vibrating granules.

The experimental setup of Fig. A1 consists of elastic granular beads and are driven harmonically. The soft plastic material acting as a vise jaw applies a precompression that reduces vibrational transmissions. The external driving frequency $(\omega_D)$ is swept from a lower value of 2 kHz to a higher value of 8 kHz with an increment of 100 Hz at a fixed voltage amplitude of $120 V_{p-p}$. The nonlinear granular beads (52100 Alloy Steel: McMaster-Carr 9528K33, 1-inch diameter, Young's Modulus 210 GPa, density 7810 kg/m³) are adaptable to different frequencies and amplitudes. The system is driven by a single transducer (V133-RM Olympus IMS) at one end. The transducer is driven through an amplifier (PD200 60W high bandwidth, low noise linear amplifier) and coupled with a waveform generator (B&K Precision 4055B). Three similar transducers are used to record responses. One record transducer is placed in the longitudinal direction, and two record transducers are placed in the lateral direction. The signal responses from the three recording transducers are measured through the oscilloscope (Tektronix MDO3024 Mixed Domain Oscilloscope) and averaged across 512-time series to minimize recording error. The generation of the varying frequency and amplitude and the recording of the signals and data are controlled and processed by MATLAB-based custom algorithms. Uniform compression force $F_0$ is provided at both ends of the system using a bench vise to fix the initial displacement $\delta_0$ between the granules centers. The responses are obtained from the lateral detecting transducers of the granules at a steady state. The alignment of the granules is center to center and guaranteed to the limit of human eye precision. Mechanical disturbance can occur in the system due to the natural frequency of the vise holding the setup. We processed the data for the sampling frequency of 999.89 kHz, and considering the noises occurring from the environment, the data were filtered by a low pass frequency of 175 kHz. The resultant signal is processed till 25 kHz to get the pure response of the signal data. This can be implemented to get the current system's signal-to-noise ratio (SNR). The signal-to-noise ratio (SNR) is a measure used to quantify the quality of a signal in the presence of noise [43]. In acoustics, it represents the ratio of the signal's power to the noise's power.